\documentclass[reprint,
 amsmath,amssymb
 aps,
 prl,
 superscriptaddress
]{revtex4-1}
\usepackage[utf8]{inputenc}
\usepackage{graphicx}% Include figure files
\usepackage{dcolumn}% Align table columns on decimal point
\usepackage{bm}% bold math
\usepackage{color}

\begin{document}
\title{Surface Elastic Constants of a Soft Solid}
\author{Qin Xu}
\affiliation{Department of Materials, ETH Z\"urich, 8093 Z\"urich, Switzerland.}

\author{Robert W. Style} 
\affiliation{Department of Materials, ETH Z\"urich, 8093 Z\"urich, Switzerland.}

\author{Eric R. Dufresne}
\email{eric.dufresne@mat.ethz.ch}
\affiliation{Department of Materials, ETH Z\"urich, 8093 Z\"urich, Switzerland.}

\begin{abstract}
Solid interfaces have intrinsic elasticity. However in most experiments, this is obscured by bulk stresses. Through microscopic observations of the contact-line geometry of a partially wetting droplet on an anisotropically stretched substrate, we measure two surface-elastic constants that quantify the linear dependence of the surface stress of a soft polymer gel on its  strain. With these two parameters, one can predict surface stresses for general deformations of the material in the linear-elastic limit. 
\end{abstract}

\maketitle

Surface stress describes the in-plane force per unit length required to deform a material interface. 
While the concept is widely used to characterize capillarity on liquid interfaces, the surface stresses of solids are normally overlooked as they are too weak to significantly deform bulk solids.  
However, recent works have shown that solid surface stress is essential to understand contact mechanics on soft-solid interfaces at $\mu$m to nm scales \cite{Style2017,Andreotti2016,Pham2017,Chakrabarti2013,Salez2013,GAO2014566,Schulman2017,Ina2017,Cao2015}. For a solid material with Young modulus, $E$, and surface stress, $\Upsilon$, solid capillarity has been shown to dominate over bulk elasticity at scales smaller than  the  elastocapillary length scale, $l_e=\Upsilon/E$ \cite{Style2013,Duan2006,bico2017}.
In this regime, the significant force contribution from surface stress, $\Upsilon$, fundamentally alters experimental behavior, and consequently the interpretation of many materials characterization techniques.
Examples where surface stress can play an important role include atomic force microscopy probing of nanowire interfaces \cite{Cuenot2004,Jing2006},  indentation tests  \cite{Jensen2017,Pham2017,Xu2016,Vella2017} and wettability measurements of soft solids \cite{Park2014,Karpitschka2015,Weijs2013,bostwick2014,nadermann2013solid,neukirch2014,mondal2015}. 

In contrast to the surface tension of simple liquids, solid surface stresses are generally expected to be a strain-dependent tensor \cite{Shuttleworth1950,Cammarata1994}. For small surface deformations, solid surface stresses are related to a strain dependent surface energy, $\gamma$, through the Shuttleworth equation \cite{Shuttleworth1950,spaepen2000}
\begin{equation}
\Upsilon_{ij}=\gamma\delta_{ij}+\frac{\partial \gamma}{\partial\epsilon^s_{ij}}
\label{shuttle-equation}
\end{equation}
where $\delta_{ij}$ is the unit tensor and $\epsilon^s_{ij}$ is  the surface-strain tensor. 
In a recent study, we observed strain dependence of the surface stress by considering the microscropic contact line geometry of a partially wetting liquid droplet on an equibiaxially-stretched, soft, silicone gel \cite{Xu2017}.  
We measured how the radial stress at the contact line, $\Upsilon_{rr}$, has a linear dependence on the local, equibiaxial strain $\epsilon^s$, given by $\Upsilon_{rr}(\epsilon^s)=\gamma_0+\Lambda\epsilon^s$.
The surface stress increases rapidly with strain, due to the large value of the elastic constant, $\Lambda=126\,$mN/m. 
While that measurement established the strain-dependence of the gel's surface stress, a proper characterization of linear surface elasticity requires the measurement of two elastic constants \cite{Gurtin1975,Ackland1986,cammarata_eby_1991}.  
With both of these material parameters,  surface stress would be predictable for any deformation in the linear limit.

Here we probe the surface elasticity of a uniaxially-stretched, soft, silicone gel by examining the wetting ridge formed at the contact line of partially wetting glycerol droplets \cite{carre1996,white2003,pericet2008,Style2013_PRL}. 
The surface stress is found to be anisotropic, depending on the local orientation of the droplet's surface relative to the applied strain.
The strain and orientation dependence of the surface stress can be fully described through surface Lam\'e coefficients, $\lambda_s$ and $\mu_s$.
At zero strain, the surface stress is isotropic and equal to $22\,$mN/m.  
For applied strains of up to 28\%, we find that $\lambda_s$ and $\mu_s$ are constant and equal to $43\,$mN/m and $20\,$mN/m respectively.

\begin{figure}[t]
\centering
\includegraphics[width=0.49\textwidth]{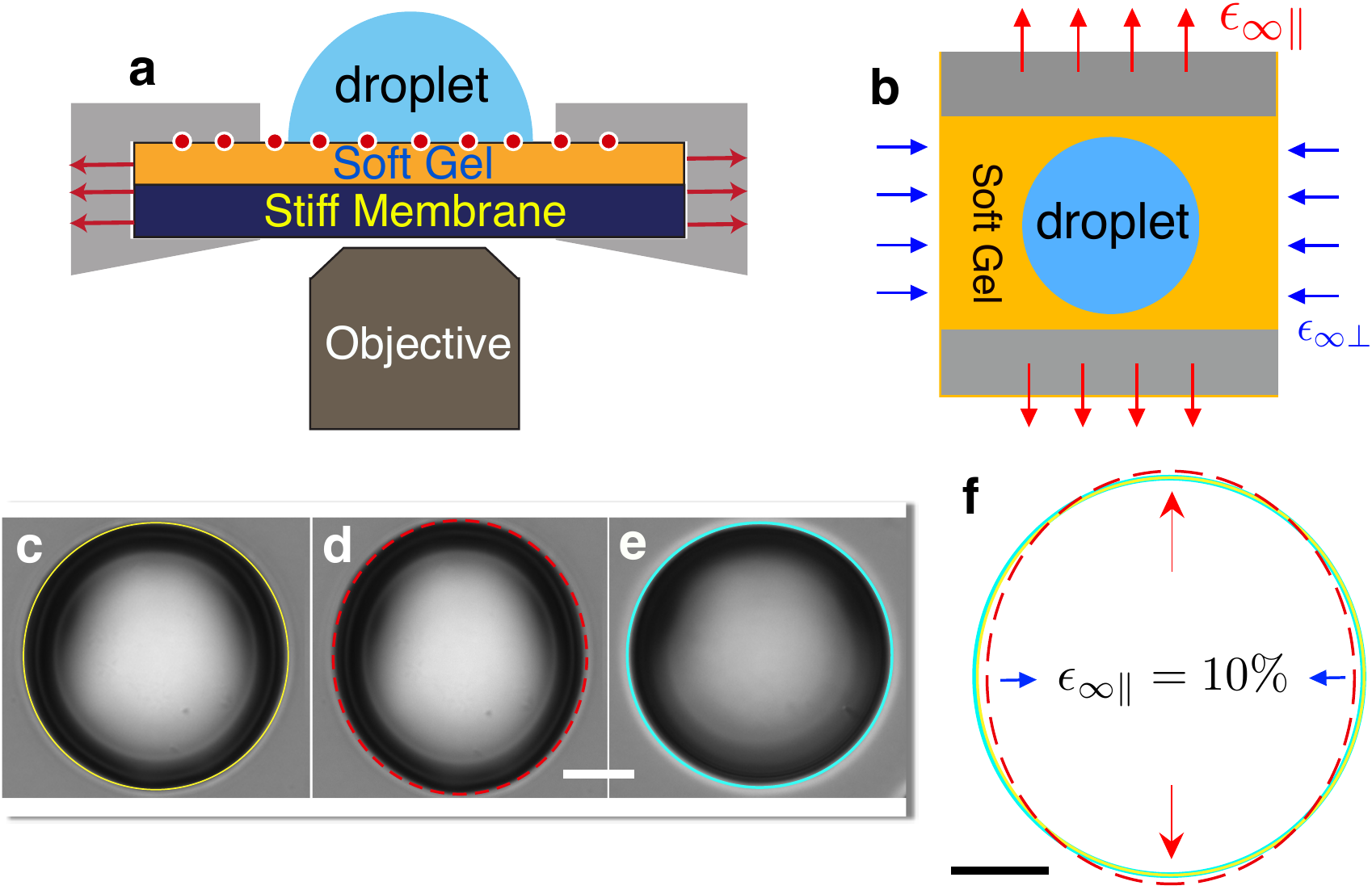}
\caption{{\bf Experimental Schematics.} (a) Side view of the experimental setup. A layer of soft gel is deposited on top of a stiffer membrane. Two sides of the membrane are then clamped to a uniaxial stretching device, and a glycerol droplet is placed on the surface. To visualize surface deformations, fluorescent beads are embedded just below the upper surface of the soft gel (indicated by the red beads), and the whole system is placed on a confocal microscope. (b) Top view. Moving the clamped sides of the membrane outwards generates a far-field strain $\epsilon_{\infty\parallel}$ along the stretching direction and a slight compressive strain field $\epsilon_{\infty\bot}$ along the perpendicular direction. (c) Optical microscopic images of a droplet right before applying stretch. (d) The droplet image right after applying a far-field stretch $\epsilon_{\infty\parallel}=10\%$  and $\epsilon_{\infty\perp}=-1.3\%$. (e) Droplet image at two hours after the stretch. (f) To directly show the relaxation of the droplet, we plot the traces of drop boundaries in panels (c) (yellow), (d) (red) and (e) (blue) together. All the error bars correspond to 200 $\mu$m.}
\end{figure}

\begin{figure*}[t]
\centering
\includegraphics[width=0.9\textwidth]{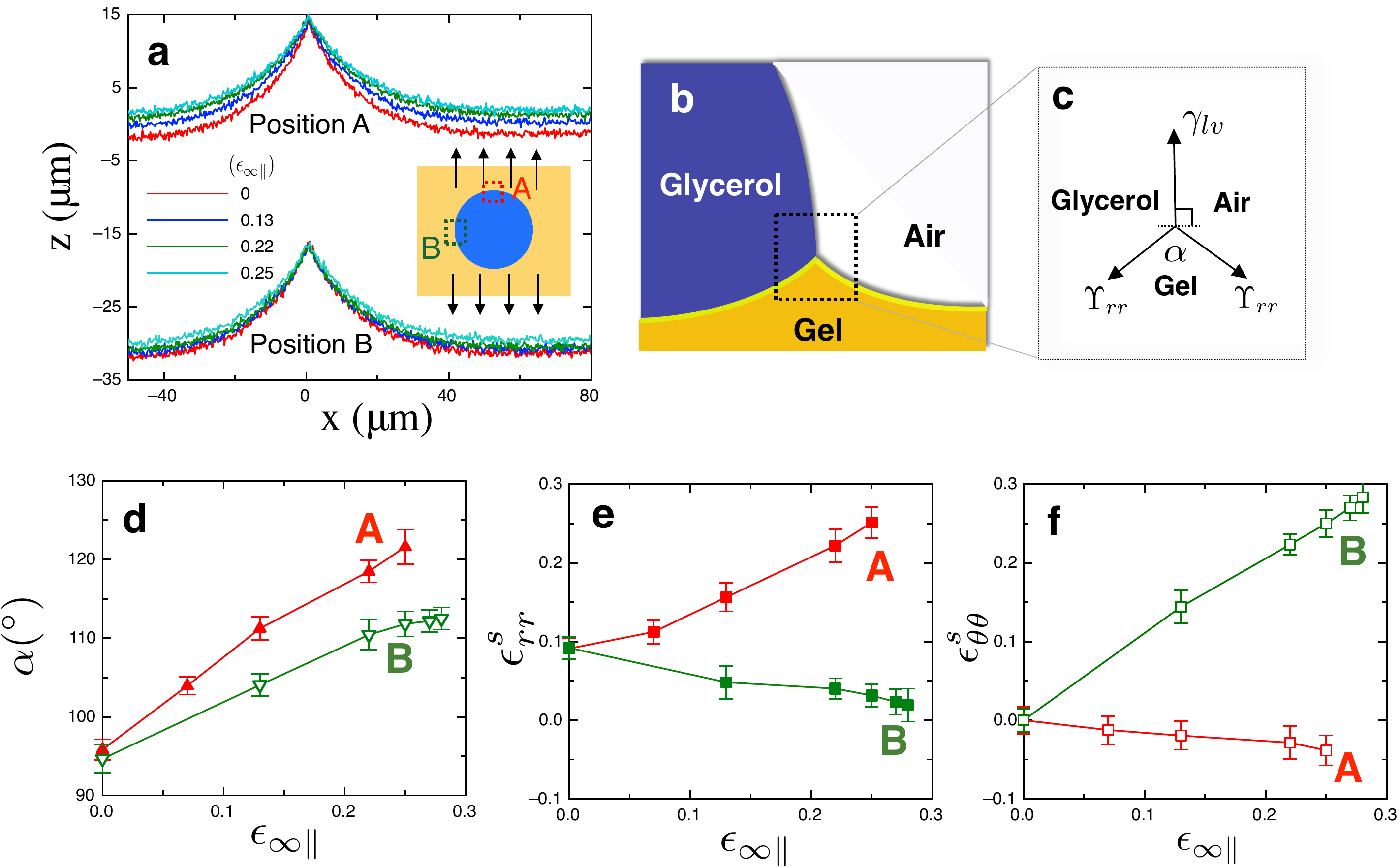}
\caption{{\bf Microscopic wetting profiles}. (a) Local wetting profiles at positions A and B (see the inset schematic) under strains of $\epsilon_{\infty\parallel}=0,0.13,0.22$ and $0.25$. At points A/B, the contact line is perpendicular/parallel to the stretch respectively.  three-phase contact line. (c) The angles between interfaces at the contact line are given by a Neumann-triangle like stress balance between the three interfacial tensions. In our experiments, the ridge is always symmetric, indicating that the solid-liquid and solid-vapor surface stresses are equal. (d) The wetting-ridge-tip angle, $\alpha$ increases with applied strain $\epsilon_{\infty}$, but faster when the applied strain is perpendicular to the contact line (point A). (e,f) Plots of the measured local strains, $\epsilon_{rr}$ and $\epsilon_{\theta\theta}$, at the wetting-ridge tip as a function of the applied strain $\epsilon_{\infty\parallel}$. These include contributions both from the applied stretch, and from deformations that arise during the growth of the wetting ridge.}
\end{figure*}

Soft silicone (Dow Corning CY52-276) substrates were prepared as described previously \cite{Xu2016}.
The film is 80 $\mu$m thick, has Young modulus, $E= 3.0\,$kPa, Poisson ratio, $\nu=0.496$, and is supported by an  stiff silicone membrane (SMI). 
The membrane was clamped to a uniaxial stretcher \cite{schulman2015} at two opposite edges that have a separation that can be finely controlled with a micro-meter stage (see Fig. 1a). 
We use the stretcher to apply a far-field strain field, $\epsilon_{\infty\parallel}$, which simultaneously results in a smaller, compressive strain, $\epsilon_{\infty\perp}$, in the perpendicular direction (see Fig. 1b).  
These are quantified by locating and tracking the displacement of fluorescent beads (Life Technologies, F-8795) embedded just under the surface of the silicone gel \cite{Boltyanskiy2017}.

To avoid any effects of contact-line hysteresis, we developed a standard protocol for applying the droplet to the substrate.
We first deposited a glycerol droplet with a radius around $500$ $\mu$m at the center of the unstretched substrate. 
After waiting for 40 minutes for the droplet to relax, we apply a strain. 
Immediately after the stretch is applied, the droplet has an elliptical shape.
However, we wait for at least two hours, as within this time, the droplet's contact line returns to a circular shape identical to the unstretched droplet (see Figs. 1c -e). This lack of pinning occurs as the contact line of glycerol can freely move along our soft gels \cite{Xu2017,Style2013_PNAS}. 
The contact line relaxation can be clearly seen in Fig. 1f which compares the droplet boundaries extracted from equilibrium state in Fig. 1c,e (yellow and light blue) to the elliptical outline right after the stretch from panel d (red). Note that the final droplet shape is also the same if it is placed on the substrate after stretching.

While the macroscopic contact angle is determined by the solid surface energy, microscopic wetting profiles near the contact line are governed by the radial component of surface stress $\Upsilon_{rr}$ \cite{olives2010,Style2013_PRL}. 
We quantify the local surface deformation near the contact line by locating particles in confocal fluorescence image stacks, as described previously in Ref. \cite{Boltyanskiy2017} .
Wetting profiles are shown at two different locations along the contact line in Fig. 2a, where the contact line is either perpendicular ({\bf A}) or parallel ({\bf B}) to the applied stretch. 
In these plots, the droplet lies to the left of the wetting ridge, which is roughly 15$\,\mu$m tall. 
While the profiles at region {\bf A} exhibit clear dependence on stretch, the change of the profiles at {\bf B} is more subtle. 
The local ridge geometry is determined by a local balance of the surface stresses \cite{Style2013_PRL} (see Fig 2c).
Since the macroscopic contact angle always remains close to 90$^\circ$, the ridge symmetry suggests that the solid-liquid and solid-vapor surface stresses are equal.
Applying a force balance in the out-of-plane direction, we can thus equate the liquid surface tension $\gamma_{lv}$ with the vertical component of the two solid surface stresses: 
\begin{equation}
\Upsilon_{rr}=\frac{\gamma_{lv}}{2\cos{(\alpha/2)}}.
\end{equation}
Here, $\alpha$ is the opening angle of the wedge \cite{Xu2016}. 
In the unstretched state,  $\alpha$ is the same around the circumference of the droplet  ($\alpha\vert_A=95.6 \pm 1.3^\circ$ and $\alpha\vert_B=94.7\pm 1.8^\circ$). 
However, as $\epsilon_{\infty\parallel}$ increases, $\alpha$ changes at points  {\bf A} and {\bf B} quite differently.
This is demonstrated in Fig. 2d.
$\alpha$ is much more sensitive to applied strain when  the contact line is perpendicular, rather than  parallel, to the applied strain.
Thus, Eq. 2 suggests that the surface stress increases faster along the direction of applied stretch than in the orthogonal direction as $\epsilon_{\infty\parallel}$ increases.

\begin{figure}[t]
\centering
\includegraphics[width=0.46\textwidth]{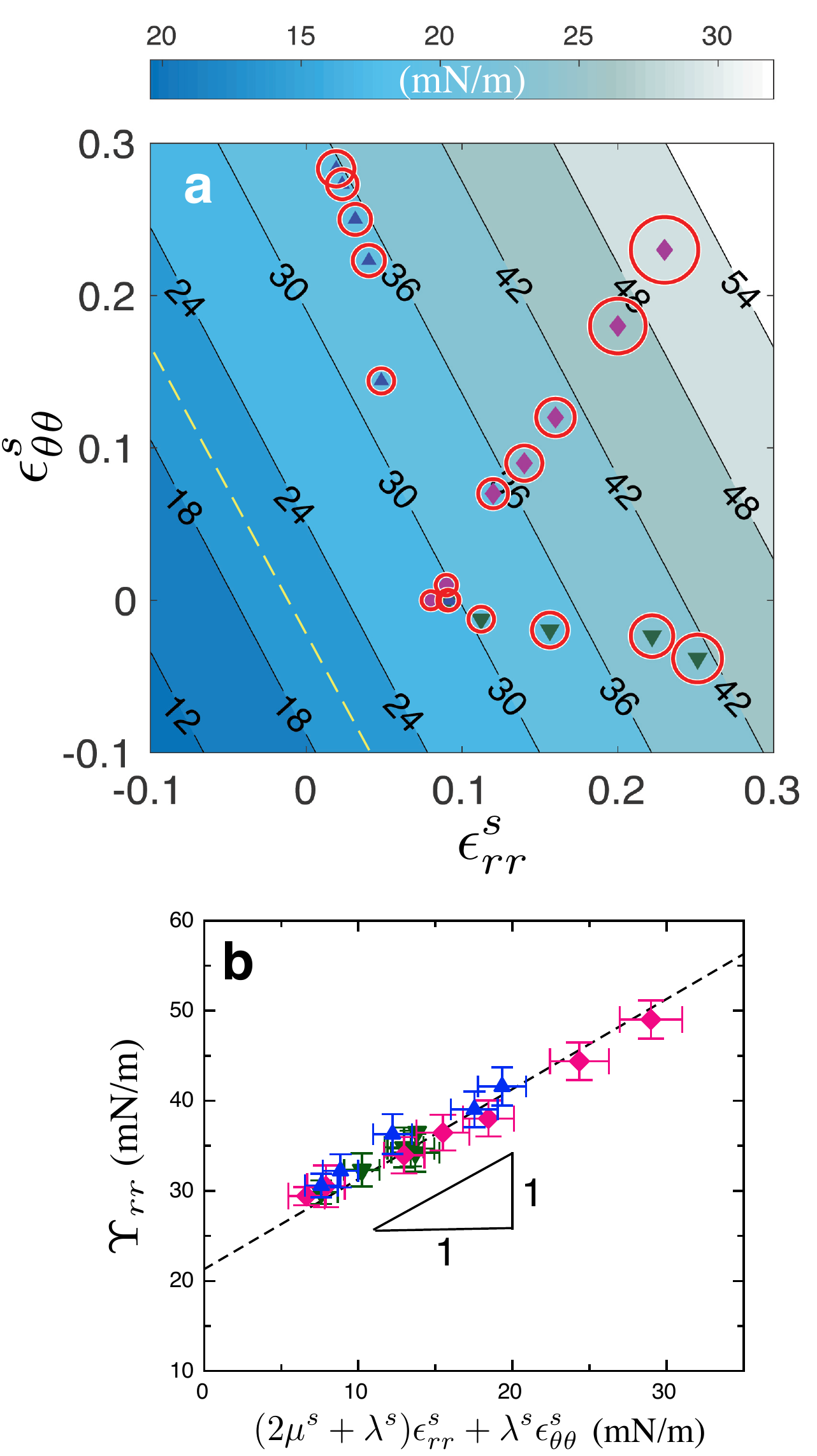}
\caption{{\bf Fitting the surface Lam\'e coefficients.} (a) The dependence of surface stress $\Upsilon_{rr}$ on $\epsilon_{rr}$ and $\epsilon_{\theta\theta}$.
For each data point the radius of the circle indicates the measured value of the surface stress. 
The underlying colormap shows the prediction of Eq. 5 using the best-fit values of $\lambda_s$ and $\mu_s$.
Here blue uptriangles and green downtriangles correspond to the experimental data points from points A and B, respectively. The pink diamonds are the biaxial stretcher results from \cite{Xu2017}. The yellow dashed line indicates the zero-strain surface stress $\Upsilon_0$.
(b) Measured values of $\Upsilon_{rr}$ agree very well with the predictions of Eq. 5, using the best fit-values of $\lambda_s$ and $\mu_s$.}

\end{figure}

To understand the origin of the anisotropy in surface stress, we need to relate local changes in the surface stress to local surface strain.
We use $\hat{r}$ and $\hat\theta$ to symbolize directions normal and tangent  to the contact line, respectively.
Accordingly, $\epsilon^s_{rr}$ and $\epsilon^s_{\theta\theta}$ correspond to the local, normal surface strains in the radial and hoop directions, while $\epsilon^s_{r\theta}$  is the shear strain. 
The locations of {\bf A} and {\bf B}  are chosen such that $\epsilon^s_{r\theta}=0$.
We measure $\epsilon^s_{rr}$ and $\epsilon^s_{\theta\theta}$ by combining the strains that arise from two sources: the applied strain from the stretcher, and further surface stretching due to the formation of the wetting ridge. 
First, we measure the far-field strains ($\epsilon_{\infty\parallel},\epsilon_{\infty\perp}$) from visualising sheet displacements before addition of a droplet. 
Second, we determine the additional local strains caused by wetting ridge formation.
The latter can be obtained by tracking the movements of fluorescent beads in 3D after placing the droplets (see \cite{Xu2016} and the Supplemental Information).  
The variation of the local strain with the far field strain at {\bf A} and {\bf B} are shown in Fig. 2e-f. 
At {\bf A},  the radial strain $\epsilon^s_{rr}$ increases significantly with $\epsilon_{\infty\parallel}$ while $\epsilon^s_{\theta\theta}$ slightly decreases due to the contractive strain $\epsilon_{\infty\perp}$.
On the other hand, $\epsilon^s_{rr}$ and $\epsilon^s_{\theta\theta}$ show the opposite trends at {\bf B}.

The variation of the surface stress with the local strain field is shown in Fig. 3a.
The surface stress increases with both $\epsilon^s_{rr}$ and $\epsilon^s_{\theta\theta}$, as shown by the size of the circle around each data point.  
We combine  uniaxial stretch measurements  at  points {\bf A} and {\bf B} (blue up triangles and green down triangles respectively) with previously reported measurements for equibiaxial stretch on the same material (shown in pink diamonds) \cite{Xu2017}.

To interpret the measured strain-dependence of the surface stress, we consider the relation between the surface-stress tensor $\Upsilon_{ij}(\epsilon^s_{ij})$ and surface energy $\gamma(\epsilon^s_{ij})$ via the Shuttleworth equation (Eq.1).
At leading order, the most general equation for the dependence of surface energy on surface strain is

\begin{figure}[t]
\centering
\includegraphics[width=0.35\textwidth]{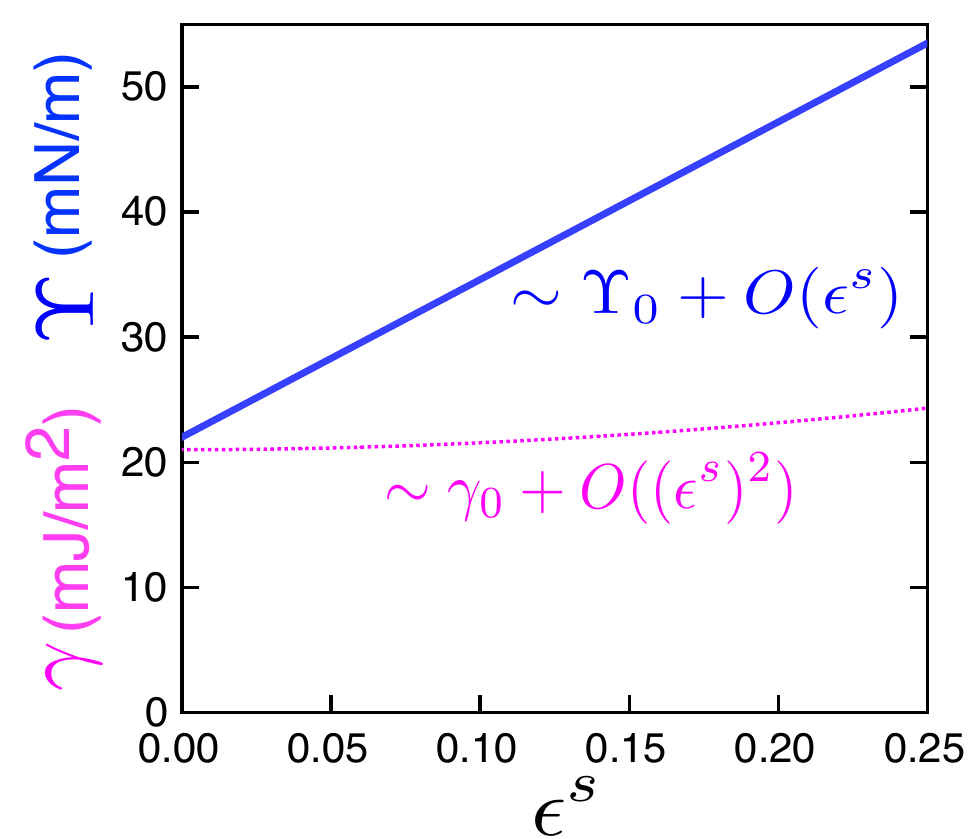}
\caption{{\bf Strain-dependent surface stress and surface energy.} Using measured elastic constant $\lambda_s$ and $\mu_s$, Eq. \ref{eq:surfaceenergy} and Eq. \ref{eq:surfacestress_normal} are plotted to show the analytic solution of surface energy (pink dashed line) and surface stress (blue solid line) for equibiaxial strain.} 
\end{figure}

\begin{equation} \label{eq:surfaceenergy} 
\gamma=\gamma_0+B_{ij}\epsilon^s_{ij}+\frac{1}{2}C_{ijkl}\epsilon^s_{ij}\epsilon^s_{kl}
\end{equation}
If we further assume that the material is isotropic, then we can write $B_{ij}=(\Upsilon_0-\gamma_0)\delta_{ij}$ and $C_{ijkl}=(\lambda^s-(\Upsilon_0-\gamma_0))\delta_{ij}\delta_{kl}+\mu^s(\delta_{ik}\delta_{jl}+\delta_{il}\delta_{jk})$.
Thus,
\begin{equation} \label{eq:surfacestress}
\Upsilon_{ij}=\Upsilon_0+\lambda^s\epsilon^s_{kk}\delta_{ij}+2\mu^s\epsilon^s_{ij},
\end{equation}
where  $\lambda^s$ and $\mu^s$ are the two surface Lam\'e coefficients. 
Since $\epsilon^s_{r\theta}=0$ at {\bf A} and {\bf B},  we find that
\begin{equation} \label{eq:surfacestress_normal}
\Upsilon_{rr}=\Upsilon_0+(2\mu^s+\lambda^s)\epsilon^s_{rr}+\lambda^s\epsilon^s_{\theta\theta}.
\end{equation}
We use this expression to fit our data for $\Upsilon_{rr} (\epsilon^s_{rr},\epsilon^s_{\theta\theta})$ to Eq. 5.
We obtain a fit with $\Upsilon_0=21.8 \pm 3.0\,$mN/m, $\lambda^s=43.1\pm 5.2\,$mN/m and $\mu^s=20.2 \pm 3.1\,$mN/m.
The accuracy of the linear-elastic model is highlighted in Fig. 3b by plotting all of the measured values of $\Upsilon_{rr}$ against the prediction of Eq.5 with the fitted Lam\'e constants.

Why do we see such a clear effect of strain on the microscopic contact line geometry,  even though  macroscopic contact angles appear to be unchanged by substrate deformation?
First, while the microscopic contact line geometry depends on the surface stresses, the macroscopic contact angle depends on the surface energies through Young's law, $\gamma_{lv} \cos \theta = \gamma_{sv}-\gamma_{sl}$ \cite{Style2013}.
Since we find that the zero-strain surface stress is indistinguishable from the expected surface energy of silicone polymers ($\Upsilon_0\approx \gamma_0$, cf \cite{hui2013}), the linear term in Eq. 3 disappears.
Thus, changes in surface energy, $\Delta \gamma$, scale as $(\epsilon^s)^2$. 
On the other hand, we find that changes in surface stress, $\Delta \Upsilon$, vary as $\epsilon^s$. 
Therefore, we generally expect $\Delta \Upsilon \gg \Delta \gamma$ for small strains. 
This is highlighted in Fig. 4, where we plot $\gamma$ and $\Upsilon_{rr}$ against equibiaxial strain using measured values of
$\lambda^s$ and $\mu^s$.
Even for rather large strains of 25\%, changes in $\Upsilon$ are much bigger than changes in $\gamma$.
Second, the macroscopic contact angle depends on the \emph{difference} of the solid surface energies against the liquid and vapor phases.
Thus, changes to the macroscopic contact angle can only occur when the surface elastic constants are different on either side of the contact line.
Together, these observations demonstrate why the microscopic line geometry is more sensitive to strain than the classic macroscopic contact angle. 

Our results provide a full description of the surface elasticity of a polymer gel.
While surface elastic constants have be widely investigated for complex fluid-fluid interfaces \cite{Fuller2012,Pepicelli2017}, we are not aware of any measurement of the surface elastic constants of any solid material.
Theoretical  calculations have predicted the elastic constants for hard materials\cite{Shenoy2005}, but we are aware of no theoretical predictions for soft solids. An understanding of the relationship between the structure and properties of the surface will be an important topic for future investigation.
Rationale control of the interfacial structure could enable new mechanical properties for applications in wetting, adhesion, and surface instabilities.

We thank Thomas Salez, Matteo Ciccotti, and Jan Vermant for useful discussions. We thank Hanumantha Rao Vutukuri and Gabriele Colombo for helping the confocal imaging.

%\bibliography{QX_citation.bib}

\begin{thebibliography}{43}%
\makeatletter
\providecommand \@ifxundefined [1]{%
 \@ifx{#1\undefined}
}%
\providecommand \@ifnum [1]{%
 \ifnum #1\expandafter \@firstoftwo
 \else \expandafter \@secondoftwo
 \fi
}%
\providecommand \@ifx [1]{%
 \ifx #1\expandafter \@firstoftwo
 \else \expandafter \@secondoftwo
 \fi
}%
\providecommand \natexlab [1]{#1}%
\providecommand \enquote  [1]{``#1''}%
\providecommand \bibnamefont  [1]{#1}%
\providecommand \bibfnamefont [1]{#1}%
\providecommand \citenamefont [1]{#1}%
\providecommand \href@noop [0]{\@secondoftwo}%
\providecommand \href [0]{\begingroup \@sanitize@url \@href}%
\providecommand \@href[1]{\@@startlink{#1}\@@href}%
\providecommand \@@href[1]{\endgroup#1\@@endlink}%
\providecommand \@sanitize@url [0]{\catcode `\\12\catcode `\$12\catcode
  `\&12\catcode `\#12\catcode `\^12\catcode `\_12\catcode `\%12\relax}%
\providecommand \@@startlink[1]{}%
\providecommand \@@endlink[0]{}%
\providecommand \url  [0]{\begingroup\@sanitize@url \@url }%
\providecommand \@url [1]{\endgroup\@href {#1}{\urlprefix }}%
\providecommand \urlprefix  [0]{URL }%
\providecommand \Eprint [0]{\href }%
\providecommand \doibase [0]{http://dx.doi.org/}%
\providecommand \selectlanguage [0]{\@gobble}%
\providecommand \bibinfo  [0]{\@secondoftwo}%
\providecommand \bibfield  [0]{\@secondoftwo}%
\providecommand \translation [1]{[#1]}%
\providecommand \BibitemOpen [0]{}%
\providecommand \bibitemStop [0]{}%
\providecommand \bibitemNoStop [0]{.\EOS\space}%
\providecommand \EOS [0]{\spacefactor3000\relax}%
\providecommand \BibitemShut  [1]{\csname bibitem#1\endcsname}%
\let\auto@bib@innerbib\@empty
%</preamble>
\bibitem [{\citenamefont {Style}\ \emph {et~al.}(2017)\citenamefont {Style},
  \citenamefont {Jagota}, \citenamefont {Hui},\ and\ \citenamefont
  {Dufresne}}]{Style2017}%
  \BibitemOpen
  \bibfield  {author} {\bibinfo {author} {\bibfnamefont {R.~W.}\ \bibnamefont
  {Style}}, \bibinfo {author} {\bibfnamefont {A.}~\bibnamefont {Jagota}},
  \bibinfo {author} {\bibfnamefont {C.-Y.}\ \bibnamefont {Hui}}, \ and\
  \bibinfo {author} {\bibfnamefont {E.~R.}\ \bibnamefont {Dufresne}},\ }\href
  {\doibase 10.1146/annurev-conmatphys-031016-025326} {\bibfield  {journal}
  {\bibinfo  {journal} {Annual Review of Condensed Matter Physics}\ }\textbf
  {\bibinfo {volume} {8}},\ \bibinfo {pages} {99} (\bibinfo {year}
  {2017})}\BibitemShut {NoStop}%
\bibitem [{\citenamefont {Andreotti}\ \emph {et~al.}(2016)\citenamefont
  {Andreotti}, \citenamefont {B{\"{a}}umchen}, \citenamefont {Boulogne},
  \citenamefont {Daniels}, \citenamefont {Dufresne}, \citenamefont {Perrin},
  \citenamefont {Salez}, \citenamefont {Snoeijer},\ and\ \citenamefont
  {Style}}]{Andreotti2016}%
  \BibitemOpen
  \bibfield  {author} {\bibinfo {author} {\bibfnamefont {B.}~\bibnamefont
  {Andreotti}}, \bibinfo {author} {\bibfnamefont {O.}~\bibnamefont
  {B{\"{a}}umchen}}, \bibinfo {author} {\bibfnamefont {F.}~\bibnamefont
  {Boulogne}}, \bibinfo {author} {\bibfnamefont {K.~E.}\ \bibnamefont
  {Daniels}}, \bibinfo {author} {\bibfnamefont {E.~R.}\ \bibnamefont
  {Dufresne}}, \bibinfo {author} {\bibfnamefont {H.}~\bibnamefont {Perrin}},
  \bibinfo {author} {\bibfnamefont {T.}~\bibnamefont {Salez}}, \bibinfo
  {author} {\bibfnamefont {J.~H.}\ \bibnamefont {Snoeijer}}, \ and\ \bibinfo
  {author} {\bibfnamefont {R.~W.}\ \bibnamefont {Style}},\ }\href {\doibase
  10.1039/C5SM03140K} {\bibfield  {journal} {\bibinfo  {journal} {Soft Matter}\
  }\textbf {\bibinfo {volume} {12}},\ \bibinfo {pages} {2993} (\bibinfo {year}
  {2016})}\BibitemShut {NoStop}%
\bibitem [{\citenamefont {Pham}\ \emph {et~al.}(2017)\citenamefont {Pham},
  \citenamefont {Schellenberger}, \citenamefont {Kappl},\ and\ \citenamefont
  {Butt}}]{Pham2017}%
  \BibitemOpen
  \bibfield  {author} {\bibinfo {author} {\bibfnamefont {J.~T.}\ \bibnamefont
  {Pham}}, \bibinfo {author} {\bibfnamefont {F.}~\bibnamefont
  {Schellenberger}}, \bibinfo {author} {\bibfnamefont {M.}~\bibnamefont
  {Kappl}}, \ and\ \bibinfo {author} {\bibfnamefont {H.-J.}\ \bibnamefont
  {Butt}},\ }\href {\doibase 10.1103/PhysRevMaterials.1.015602} {\bibfield
  {journal} {\bibinfo  {journal} {Phys. Rev. Materials}\ }\textbf {\bibinfo
  {volume} {1}},\ \bibinfo {pages} {015602} (\bibinfo {year}
  {2017})}\BibitemShut {NoStop}%
\bibitem [{\citenamefont {Chakrabarti}\ and\ \citenamefont
  {Chaudhury}(2013)}]{Chakrabarti2013}%
  \BibitemOpen
  \bibfield  {author} {\bibinfo {author} {\bibfnamefont {A.}~\bibnamefont
  {Chakrabarti}}\ and\ \bibinfo {author} {\bibfnamefont {M.~K.}\ \bibnamefont
  {Chaudhury}},\ }\href {\doibase 10.1021/la401115j} {\bibfield  {journal}
  {\bibinfo  {journal} {Langmuir}\ }\textbf {\bibinfo {volume} {29}},\ \bibinfo
  {pages} {6926} (\bibinfo {year} {2013})}\BibitemShut {NoStop}%
\bibitem [{\citenamefont {Salez}\ \emph {et~al.}(2013)\citenamefont {Salez},
  \citenamefont {Benzaquen},\ and\ \citenamefont {Rapha{\"{e}}l}}]{Salez2013}%
  \BibitemOpen
  \bibfield  {author} {\bibinfo {author} {\bibfnamefont {T.}~\bibnamefont
  {Salez}}, \bibinfo {author} {\bibfnamefont {M.}~\bibnamefont {Benzaquen}}, \
  and\ \bibinfo {author} {\bibfnamefont {{\'{E}}.}~\bibnamefont
  {Rapha{\"{e}}l}},\ }\href {http://xlink.rsc.org/?DOI=c3sm51780b} {\bibfield
  {journal} {\bibinfo  {journal} {Soft Matter}\ }\textbf {\bibinfo {volume}
  {9}},\ \bibinfo {pages} {10699} (\bibinfo {year} {2013})}\BibitemShut
  {NoStop}%
\bibitem [{\citenamefont {Gao}\ \emph {et~al.}(2014)\citenamefont {Gao},
  \citenamefont {Hao}, \citenamefont {Huang},\ and\ \citenamefont
  {Fang}}]{GAO2014566}%
  \BibitemOpen
  \bibfield  {author} {\bibinfo {author} {\bibfnamefont {X.}~\bibnamefont
  {Gao}}, \bibinfo {author} {\bibfnamefont {F.}~\bibnamefont {Hao}}, \bibinfo
  {author} {\bibfnamefont {Z.}~\bibnamefont {Huang}}, \ and\ \bibinfo {author}
  {\bibfnamefont {D.}~\bibnamefont {Fang}},\ }\href {\doibase
  https://doi.org/10.1016/j.ijsolstr.2013.10.017} {\bibfield  {journal}
  {\bibinfo  {journal} {International Journal of Solids and Structures}\
  }\textbf {\bibinfo {volume} {51}},\ \bibinfo {pages} {566 } (\bibinfo {year}
  {2014})}\BibitemShut {NoStop}%
\bibitem [{\citenamefont {Schulman}\ \emph {et~al.}(2017)\citenamefont
  {Schulman}, \citenamefont {Ledesma-Alonso}, \citenamefont {Salez},
  \citenamefont {Rapha\"el},\ and\ \citenamefont
  {Dalnoki-Veress}}]{Schulman2017}%
  \BibitemOpen
  \bibfield  {author} {\bibinfo {author} {\bibfnamefont {R.~D.}\ \bibnamefont
  {Schulman}}, \bibinfo {author} {\bibfnamefont {R.}~\bibnamefont
  {Ledesma-Alonso}}, \bibinfo {author} {\bibfnamefont {T.}~\bibnamefont
  {Salez}}, \bibinfo {author} {\bibfnamefont {E.}~\bibnamefont {Rapha\"el}}, \
  and\ \bibinfo {author} {\bibfnamefont {K.}~\bibnamefont {Dalnoki-Veress}},\
  }\href {\doibase 10.1103/PhysRevLett.118.198002} {\bibfield  {journal}
  {\bibinfo  {journal} {Phys. Rev. Lett.}\ }\textbf {\bibinfo {volume} {118}},\
  \bibinfo {pages} {198002} (\bibinfo {year} {2017})}\BibitemShut {NoStop}%
\bibitem [{\citenamefont {Ina}\ \emph {et~al.}(2017)\citenamefont {Ina},
  \citenamefont {Cao}, \citenamefont {Vatankhah-Varnoosfaderani}, \citenamefont
  {Everhart}, \citenamefont {Daniel}, \citenamefont {Dobrynin},\ and\
  \citenamefont {Sheiko}}]{Ina2017}%
  \BibitemOpen
  \bibfield  {author} {\bibinfo {author} {\bibfnamefont {M.}~\bibnamefont
  {Ina}}, \bibinfo {author} {\bibfnamefont {Z.}~\bibnamefont {Cao}}, \bibinfo
  {author} {\bibfnamefont {M.}~\bibnamefont {Vatankhah-Varnoosfaderani}},
  \bibinfo {author} {\bibfnamefont {M.~H.}\ \bibnamefont {Everhart}}, \bibinfo
  {author} {\bibfnamefont {W.~F.}\ \bibnamefont {Daniel}}, \bibinfo {author}
  {\bibfnamefont {A.~V.}\ \bibnamefont {Dobrynin}}, \ and\ \bibinfo {author}
  {\bibfnamefont {S.~S.}\ \bibnamefont {Sheiko}},\ }\href@noop {} {\bibfield
  {journal} {\bibinfo  {journal} {ACS Macro Letters}\ }\textbf {\bibinfo
  {volume} {6}},\ \bibinfo {pages} {854} (\bibinfo {year} {2017})}\BibitemShut
  {NoStop}%
\bibitem [{\citenamefont {Cao}\ and\ \citenamefont {Dobrynin}(2015)}]{Cao2015}%
  \BibitemOpen
  \bibfield  {author} {\bibinfo {author} {\bibfnamefont {Z.}~\bibnamefont
  {Cao}}\ and\ \bibinfo {author} {\bibfnamefont {A.~V.}\ \bibnamefont
  {Dobrynin}},\ }\href@noop {} {\bibfield  {journal} {\bibinfo  {journal}
  {Langmuir}\ }\textbf {\bibinfo {volume} {31}},\ \bibinfo {pages} {12520}
  (\bibinfo {year} {2015})}\BibitemShut {NoStop}%
\bibitem [{\citenamefont {Style}\ \emph
  {et~al.}(2013{\natexlab{a}})\citenamefont {Style}, \citenamefont {Hyland},
  \citenamefont {Boltyanskiy}, \citenamefont {Wettlaufer},\ and\ \citenamefont
  {Dufresne}}]{Style2013}%
  \BibitemOpen
  \bibfield  {author} {\bibinfo {author} {\bibfnamefont {R.~W.}\ \bibnamefont
  {Style}}, \bibinfo {author} {\bibfnamefont {C.}~\bibnamefont {Hyland}},
  \bibinfo {author} {\bibfnamefont {R.}~\bibnamefont {Boltyanskiy}}, \bibinfo
  {author} {\bibfnamefont {J.~S.}\ \bibnamefont {Wettlaufer}}, \ and\ \bibinfo
  {author} {\bibfnamefont {E.~R.}\ \bibnamefont {Dufresne}},\ }\href {\doibase
  10.1038/ncomms3728} {\bibfield  {journal} {\bibinfo  {journal} {Nature
  communications}\ }\textbf {\bibinfo {volume} {4}},\ \bibinfo {pages} {2728}
  (\bibinfo {year} {2013}{\natexlab{a}})}\BibitemShut {NoStop}%
\bibitem [{\citenamefont {Duan}\ \emph {et~al.}(2005)\citenamefont {Duan},
  \citenamefont {Wang}, \citenamefont {Huang},\ and\ \citenamefont
  {Karihaloo}}]{Duan2006}%
  \BibitemOpen
  \bibfield  {author} {\bibinfo {author} {\bibfnamefont {H.}~\bibnamefont
  {Duan}}, \bibinfo {author} {\bibfnamefont {J.}~\bibnamefont {Wang}}, \bibinfo
  {author} {\bibfnamefont {Z.}~\bibnamefont {Huang}}, \ and\ \bibinfo {author}
  {\bibfnamefont {B.}~\bibnamefont {Karihaloo}},\ }\href {\doibase
  https://doi.org/10.1016/j.jmps.2005.02.009} {\bibfield  {journal} {\bibinfo
  {journal} {Journal of the Mechanics and Physics of Solids}\ }\textbf
  {\bibinfo {volume} {53}},\ \bibinfo {pages} {1574 } (\bibinfo {year}
  {2005})}\BibitemShut {NoStop}%
\bibitem [{\citenamefont {Bico}\ \emph {et~al.}(2017)\citenamefont {Bico},
  \citenamefont {Reyssat},\ and\ \citenamefont {Roman}}]{bico2017}%
  \BibitemOpen
  \bibfield  {author} {\bibinfo {author} {\bibfnamefont {J.}~\bibnamefont
  {Bico}}, \bibinfo {author} {\bibfnamefont {{\'E}.}~\bibnamefont {Reyssat}}, \
  and\ \bibinfo {author} {\bibfnamefont {B.}~\bibnamefont {Roman}},\
  }\href@noop {} {\bibfield  {journal} {\bibinfo  {journal} {Ann Rev. Fluid
  Mech.}\ } (\bibinfo {year} {2017})}\BibitemShut {NoStop}%
\bibitem [{\citenamefont {Cuenot}\ \emph {et~al.}(2004)\citenamefont {Cuenot},
  \citenamefont {Fr\'etigny}, \citenamefont {Demoustier-Champagne},\ and\
  \citenamefont {Nysten}}]{Cuenot2004}%
  \BibitemOpen
  \bibfield  {author} {\bibinfo {author} {\bibfnamefont {S.}~\bibnamefont
  {Cuenot}}, \bibinfo {author} {\bibfnamefont {C.}~\bibnamefont {Fr\'etigny}},
  \bibinfo {author} {\bibfnamefont {S.}~\bibnamefont {Demoustier-Champagne}}, \
  and\ \bibinfo {author} {\bibfnamefont {B.}~\bibnamefont {Nysten}},\ }\href
  {\doibase 10.1103/PhysRevB.69.165410} {\bibfield  {journal} {\bibinfo
  {journal} {Phys. Rev. B}\ }\textbf {\bibinfo {volume} {69}},\ \bibinfo
  {pages} {165410} (\bibinfo {year} {2004})}\BibitemShut {NoStop}%
\bibitem [{\citenamefont {Jing}\ \emph {et~al.}(2006)\citenamefont {Jing},
  \citenamefont {Duan}, \citenamefont {Sun}, \citenamefont {Zhang},
  \citenamefont {Xu}, \citenamefont {Li}, \citenamefont {Wang},\ and\
  \citenamefont {Yu}}]{Jing2006}%
  \BibitemOpen
  \bibfield  {author} {\bibinfo {author} {\bibfnamefont {G.~Y.}\ \bibnamefont
  {Jing}}, \bibinfo {author} {\bibfnamefont {H.~L.}\ \bibnamefont {Duan}},
  \bibinfo {author} {\bibfnamefont {X.~M.}\ \bibnamefont {Sun}}, \bibinfo
  {author} {\bibfnamefont {Z.~S.}\ \bibnamefont {Zhang}}, \bibinfo {author}
  {\bibfnamefont {J.}~\bibnamefont {Xu}}, \bibinfo {author} {\bibfnamefont
  {Y.~D.}\ \bibnamefont {Li}}, \bibinfo {author} {\bibfnamefont {J.~X.}\
  \bibnamefont {Wang}}, \ and\ \bibinfo {author} {\bibfnamefont {D.~P.}\
  \bibnamefont {Yu}},\ }\href {\doibase 10.1103/PhysRevB.73.235409} {\bibfield
  {journal} {\bibinfo  {journal} {Phys. Rev. B}\ }\textbf {\bibinfo {volume}
  {73}},\ \bibinfo {pages} {235409} (\bibinfo {year} {2006})}\BibitemShut
  {NoStop}%
\bibitem [{\citenamefont {Jensen}\ \emph {et~al.}(2017)\citenamefont {Jensen},
  \citenamefont {Style}, \citenamefont {Xu},\ and\ \citenamefont
  {Dufresne}}]{Jensen2017}%
  \BibitemOpen
  \bibfield  {author} {\bibinfo {author} {\bibfnamefont {K.~E.}\ \bibnamefont
  {Jensen}}, \bibinfo {author} {\bibfnamefont {R.~W.}\ \bibnamefont {Style}},
  \bibinfo {author} {\bibfnamefont {Q.}~\bibnamefont {Xu}}, \ and\ \bibinfo
  {author} {\bibfnamefont {E.~R.}\ \bibnamefont {Dufresne}},\ }\href {\doibase
  10.1103/PhysRevX.7.041031} {\bibfield  {journal} {\bibinfo  {journal} {Phys.
  Rev. X}\ }\textbf {\bibinfo {volume} {7}},\ \bibinfo {pages} {041031}
  (\bibinfo {year} {2017})}\BibitemShut {NoStop}%
\bibitem [{\citenamefont {Xu}\ \emph {et~al.}(2016)\citenamefont {Xu},
  \citenamefont {Jagota}, \citenamefont {Paretkar},\ and\ \citenamefont
  {Hui}}]{Xu2016}%
  \BibitemOpen
  \bibfield  {author} {\bibinfo {author} {\bibfnamefont {X.}~\bibnamefont
  {Xu}}, \bibinfo {author} {\bibfnamefont {A.}~\bibnamefont {Jagota}}, \bibinfo
  {author} {\bibfnamefont {D.}~\bibnamefont {Paretkar}}, \ and\ \bibinfo
  {author} {\bibfnamefont {C.-Y.}\ \bibnamefont {Hui}},\ }\href {\doibase
  10.1039/C6SM00584E} {\bibfield  {journal} {\bibinfo  {journal} {Soft Matter}\
  }\textbf {\bibinfo {volume} {12}},\ \bibinfo {pages} {5121} (\bibinfo {year}
  {2016})}\BibitemShut {NoStop}%
\bibitem [{\citenamefont {Vella}\ and\ \citenamefont
  {Davidovitch}(2017)}]{Vella2017}%
  \BibitemOpen
  \bibfield  {author} {\bibinfo {author} {\bibfnamefont {D.}~\bibnamefont
  {Vella}}\ and\ \bibinfo {author} {\bibfnamefont {B.}~\bibnamefont
  {Davidovitch}},\ }\href {http://xlink.rsc.org/?DOI=C6SM02451C} {\bibfield
  {journal} {\bibinfo  {journal} {Soft Matter}\ }\textbf {\bibinfo {volume}
  {13}},\ \bibinfo {pages} {2264} (\bibinfo {year} {2017})}\BibitemShut
  {NoStop}%
\bibitem [{\citenamefont {Park}\ \emph {et~al.}(2014)\citenamefont {Park},
  \citenamefont {Weon}, \citenamefont {Lee}, \citenamefont {Lee}, \citenamefont
  {Kim},\ and\ \citenamefont {Je}}]{Park2014}%
  \BibitemOpen
  \bibfield  {author} {\bibinfo {author} {\bibfnamefont {S.~J.}\ \bibnamefont
  {Park}}, \bibinfo {author} {\bibfnamefont {B.~M.}\ \bibnamefont {Weon}},
  \bibinfo {author} {\bibfnamefont {J.~S.}\ \bibnamefont {Lee}}, \bibinfo
  {author} {\bibfnamefont {J.}~\bibnamefont {Lee}}, \bibinfo {author}
  {\bibfnamefont {J.}~\bibnamefont {Kim}}, \ and\ \bibinfo {author}
  {\bibfnamefont {J.~H.}\ \bibnamefont {Je}},\ }\href {\doibase
  10.1038/ncomms5369} {\bibfield  {journal} {\bibinfo  {journal} {Nat.
  Commun.}\ }\textbf {\bibinfo {volume} {5}},\ \bibinfo {pages} {4369}
  (\bibinfo {year} {2014})}\BibitemShut {NoStop}%
\bibitem [{\citenamefont {Karpitschka}\ \emph {et~al.}(2015)\citenamefont
  {Karpitschka}, \citenamefont {Das}, \citenamefont {van Gorcum}, \citenamefont
  {Perrin}, \citenamefont {Andreotti},\ and\ \citenamefont
  {Snoeijer}}]{Karpitschka2015}%
  \BibitemOpen
  \bibfield  {author} {\bibinfo {author} {\bibfnamefont {S.}~\bibnamefont
  {Karpitschka}}, \bibinfo {author} {\bibfnamefont {S.}~\bibnamefont {Das}},
  \bibinfo {author} {\bibfnamefont {M.}~\bibnamefont {van Gorcum}}, \bibinfo
  {author} {\bibfnamefont {H.}~\bibnamefont {Perrin}}, \bibinfo {author}
  {\bibfnamefont {B.}~\bibnamefont {Andreotti}}, \ and\ \bibinfo {author}
  {\bibfnamefont {J.~H.}\ \bibnamefont {Snoeijer}},\ }\href {\doibase
  10.1038/ncomms8891} {\bibfield  {journal} {\bibinfo  {journal} {Nature
  Communications}\ }\textbf {\bibinfo {volume} {6}},\ \bibinfo {pages} {7891}
  (\bibinfo {year} {2015})}\BibitemShut {NoStop}%
\bibitem [{\citenamefont {Weijs}\ \emph {et~al.}(2013)\citenamefont {Weijs},
  \citenamefont {Andreotti},\ and\ \citenamefont {Snoeijer}}]{Weijs2013}%
  \BibitemOpen
  \bibfield  {author} {\bibinfo {author} {\bibfnamefont {J.~H.}\ \bibnamefont
  {Weijs}}, \bibinfo {author} {\bibfnamefont {B.}~\bibnamefont {Andreotti}}, \
  and\ \bibinfo {author} {\bibfnamefont {J.~H.}\ \bibnamefont {Snoeijer}},\
  }\href {\doibase 10.1039/c3sm50861g} {\bibfield  {journal} {\bibinfo
  {journal} {Soft Matter}\ }\textbf {\bibinfo {volume} {9}},\ \bibinfo {pages}
  {8494} (\bibinfo {year} {2013})}\BibitemShut {NoStop}%
\bibitem [{\citenamefont {Bostwick}\ \emph {et~al.}(2014)\citenamefont
  {Bostwick}, \citenamefont {Shearer},\ and\ \citenamefont
  {Daniels}}]{bostwick2014}%
  \BibitemOpen
  \bibfield  {author} {\bibinfo {author} {\bibfnamefont {J.~B.}\ \bibnamefont
  {Bostwick}}, \bibinfo {author} {\bibfnamefont {M.}~\bibnamefont {Shearer}}, \
  and\ \bibinfo {author} {\bibfnamefont {K.~E.}\ \bibnamefont {Daniels}},\
  }\href@noop {} {\bibfield  {journal} {\bibinfo  {journal} {Soft Matter}\
  }\textbf {\bibinfo {volume} {10}},\ \bibinfo {pages} {7361} (\bibinfo {year}
  {2014})}\BibitemShut {NoStop}%
\bibitem [{\citenamefont {Nadermann}\ \emph {et~al.}(2013)\citenamefont
  {Nadermann}, \citenamefont {Hui},\ and\ \citenamefont
  {Jagota}}]{nadermann2013solid}%
  \BibitemOpen
  \bibfield  {author} {\bibinfo {author} {\bibfnamefont {N.}~\bibnamefont
  {Nadermann}}, \bibinfo {author} {\bibfnamefont {C.-Y.}\ \bibnamefont {Hui}},
  \ and\ \bibinfo {author} {\bibfnamefont {A.}~\bibnamefont {Jagota}},\
  }\href@noop {} {\bibfield  {journal} {\bibinfo  {journal} {Proc. Nat. Acad.
  Sci.}\ }\textbf {\bibinfo {volume} {110}},\ \bibinfo {pages} {10541}
  (\bibinfo {year} {2013})}\BibitemShut {NoStop}%
\bibitem [{\citenamefont {Neukirch}\ \emph {et~al.}(2014)\citenamefont
  {Neukirch}, \citenamefont {Antkowiak},\ and\ \citenamefont
  {Marigo}}]{neukirch2014}%
  \BibitemOpen
  \bibfield  {author} {\bibinfo {author} {\bibfnamefont {S.}~\bibnamefont
  {Neukirch}}, \bibinfo {author} {\bibfnamefont {A.}~\bibnamefont {Antkowiak}},
  \ and\ \bibinfo {author} {\bibfnamefont {J.-J.}\ \bibnamefont {Marigo}},\
  }\href@noop {} {\bibfield  {journal} {\bibinfo  {journal} {Phys. Rev. E}\
  }\textbf {\bibinfo {volume} {89}},\ \bibinfo {pages} {012401} (\bibinfo
  {year} {2014})}\BibitemShut {NoStop}%
\bibitem [{\citenamefont {Mondal}\ \emph {et~al.}(2015)\citenamefont {Mondal},
  \citenamefont {Phukan},\ and\ \citenamefont {Ghatak}}]{mondal2015}%
  \BibitemOpen
  \bibfield  {author} {\bibinfo {author} {\bibfnamefont {S.}~\bibnamefont
  {Mondal}}, \bibinfo {author} {\bibfnamefont {M.}~\bibnamefont {Phukan}}, \
  and\ \bibinfo {author} {\bibfnamefont {A.}~\bibnamefont {Ghatak}},\
  }\href@noop {} {\bibfield  {journal} {\bibinfo  {journal} {Proc. Nat. Acad.
  Sci.}\ }\textbf {\bibinfo {volume} {112}},\ \bibinfo {pages} {12563}
  (\bibinfo {year} {2015})}\BibitemShut {NoStop}%
\bibitem [{\citenamefont {Shuttleworth}(1950)}]{Shuttleworth1950}%
  \BibitemOpen
  \bibfield  {author} {\bibinfo {author} {\bibfnamefont {R.}~\bibnamefont
  {Shuttleworth}},\ }\href {\doibase 10.1088/0370-1298/63/5/302} {\bibfield
  {journal} {\bibinfo  {journal} {Proceedings of the Physical Society. Section
  A}\ }\textbf {\bibinfo {volume} {63}},\ \bibinfo {pages} {444} (\bibinfo
  {year} {1950})}\BibitemShut {NoStop}%
\bibitem [{\citenamefont {Cammarata}(1994)}]{Cammarata1994}%
  \BibitemOpen
  \bibfield  {author} {\bibinfo {author} {\bibfnamefont {R.~C.}\ \bibnamefont
  {Cammarata}},\ }\href {\doibase 10.1016/0079-6816(94)90005-1} {\bibfield
  {journal} {\bibinfo  {journal} {Progress in Surface Science}\ }\textbf
  {\bibinfo {volume} {46}},\ \bibinfo {pages} {1} (\bibinfo {year}
  {1994})}\BibitemShut {NoStop}%
\bibitem [{\citenamefont {Spaepen}(2000)}]{spaepen2000}%
  \BibitemOpen
  \bibfield  {author} {\bibinfo {author} {\bibfnamefont {F.}~\bibnamefont
  {Spaepen}},\ }\href@noop {} {\bibfield  {journal} {\bibinfo  {journal} {Acta
  Materialia}\ }\textbf {\bibinfo {volume} {48}},\ \bibinfo {pages} {31}
  (\bibinfo {year} {2000})}\BibitemShut {NoStop}%
\bibitem [{\citenamefont {Xu}\ \emph {et~al.}(2017)\citenamefont {Xu},
  \citenamefont {Jensen}, \citenamefont {Boltyanskiy}, \citenamefont {Sarfati},
  \citenamefont {Style},\ and\ \citenamefont {Dufresne}}]{Xu2017}%
  \BibitemOpen
  \bibfield  {author} {\bibinfo {author} {\bibfnamefont {Q.}~\bibnamefont
  {Xu}}, \bibinfo {author} {\bibfnamefont {K.~E.}\ \bibnamefont {Jensen}},
  \bibinfo {author} {\bibfnamefont {R.}~\bibnamefont {Boltyanskiy}}, \bibinfo
  {author} {\bibfnamefont {R.}~\bibnamefont {Sarfati}}, \bibinfo {author}
  {\bibfnamefont {R.~W.}\ \bibnamefont {Style}}, \ and\ \bibinfo {author}
  {\bibfnamefont {E.~R.}\ \bibnamefont {Dufresne}},\ }\href@noop {} {\bibfield
  {journal} {\bibinfo  {journal} {Nature Communications}\ }\textbf {\bibinfo
  {volume} {8}},\ \bibinfo {pages} {555} (\bibinfo {year} {2017})}\BibitemShut
  {NoStop}%
\bibitem [{\citenamefont {Gurtin}\ and\ \citenamefont
  {Ian~Murdoch}(1975)}]{Gurtin1975}%
  \BibitemOpen
  \bibfield  {author} {\bibinfo {author} {\bibfnamefont {M.~E.}\ \bibnamefont
  {Gurtin}}\ and\ \bibinfo {author} {\bibfnamefont {A.}~\bibnamefont
  {Ian~Murdoch}},\ }\href {\doibase 10.1007/BF00261375} {\bibfield  {journal}
  {\bibinfo  {journal} {Archive for Rational Mechanics and Analysis}\ }\textbf
  {\bibinfo {volume} {57}},\ \bibinfo {pages} {291} (\bibinfo {year}
  {1975})}\BibitemShut {NoStop}%
\bibitem [{\citenamefont {Ackland}\ and\ \citenamefont
  {Finnis}(1986)}]{Ackland1986}%
  \BibitemOpen
  \bibfield  {author} {\bibinfo {author} {\bibfnamefont {G.~J.}\ \bibnamefont
  {Ackland}}\ and\ \bibinfo {author} {\bibfnamefont {M.~W.}\ \bibnamefont
  {Finnis}},\ }\href {\doibase 10.1080/01418618608242900} {\bibfield  {journal}
  {\bibinfo  {journal} {Philosophical Magazine A}\ }\textbf {\bibinfo {volume}
  {54}},\ \bibinfo {pages} {301} (\bibinfo {year} {1986})}\BibitemShut
  {NoStop}%
\bibitem [{\citenamefont {Cammarata}\ and\ \citenamefont
  {Eby}(1991)}]{cammarata_eby_1991}%
  \BibitemOpen
  \bibfield  {author} {\bibinfo {author} {\bibfnamefont {R.}~\bibnamefont
  {Cammarata}}\ and\ \bibinfo {author} {\bibfnamefont {R.}~\bibnamefont
  {Eby}},\ }\href {\doibase 10.1557/JMR.1991.0888} {\bibfield  {journal}
  {\bibinfo  {journal} {Journal of Materials Research}\ }\textbf {\bibinfo
  {volume} {6}},\ \bibinfo {pages} {888–890} (\bibinfo {year}
  {1991})}\BibitemShut {NoStop}%
\bibitem [{\citenamefont {Carr{\'e}}\ \emph {et~al.}(1996)\citenamefont
  {Carr{\'e}}, \citenamefont {Gastel},\ and\ \citenamefont
  {Shanahan}}]{carre1996}%
  \BibitemOpen
  \bibfield  {author} {\bibinfo {author} {\bibfnamefont {A.}~\bibnamefont
  {Carr{\'e}}}, \bibinfo {author} {\bibfnamefont {J.-C.}\ \bibnamefont
  {Gastel}}, \ and\ \bibinfo {author} {\bibfnamefont {M.~E.}\ \bibnamefont
  {Shanahan}},\ }\href@noop {} {\bibfield  {journal} {\bibinfo  {journal}
  {Nature}\ }\textbf {\bibinfo {volume} {379}},\ \bibinfo {pages} {432}
  (\bibinfo {year} {1996})}\BibitemShut {NoStop}%
\bibitem [{\citenamefont {White}(2003)}]{white2003}%
  \BibitemOpen
  \bibfield  {author} {\bibinfo {author} {\bibfnamefont {L.~R.}\ \bibnamefont
  {White}},\ }\href@noop {} {\bibfield  {journal} {\bibinfo  {journal} {J.
  Coll. Interface Sci.}\ }\textbf {\bibinfo {volume} {258}},\ \bibinfo {pages}
  {82} (\bibinfo {year} {2003})}\BibitemShut {NoStop}%
\bibitem [{\citenamefont {Pericet-C{\'a}mara}\ \emph
  {et~al.}(2008)\citenamefont {Pericet-C{\'a}mara}, \citenamefont {Best},
  \citenamefont {Butt},\ and\ \citenamefont {Bonaccurso}}]{pericet2008}%
  \BibitemOpen
  \bibfield  {author} {\bibinfo {author} {\bibfnamefont {R.}~\bibnamefont
  {Pericet-C{\'a}mara}}, \bibinfo {author} {\bibfnamefont {A.}~\bibnamefont
  {Best}}, \bibinfo {author} {\bibfnamefont {H.-J.}\ \bibnamefont {Butt}}, \
  and\ \bibinfo {author} {\bibfnamefont {E.}~\bibnamefont {Bonaccurso}},\
  }\href@noop {} {\bibfield  {journal} {\bibinfo  {journal} {Langmuir}\
  }\textbf {\bibinfo {volume} {24}},\ \bibinfo {pages} {10565} (\bibinfo {year}
  {2008})}\BibitemShut {NoStop}%
\bibitem [{\citenamefont {Style}\ \emph
  {et~al.}(2013{\natexlab{b}})\citenamefont {Style}, \citenamefont
  {Boltyanskiy}, \citenamefont {Che}, \citenamefont {Wettlaufer}, \citenamefont
  {Wilen},\ and\ \citenamefont {Dufresne}}]{Style2013_PRL}%
  \BibitemOpen
  \bibfield  {author} {\bibinfo {author} {\bibfnamefont {R.~W.}\ \bibnamefont
  {Style}}, \bibinfo {author} {\bibfnamefont {R.}~\bibnamefont {Boltyanskiy}},
  \bibinfo {author} {\bibfnamefont {Y.}~\bibnamefont {Che}}, \bibinfo {author}
  {\bibfnamefont {J.~S.}\ \bibnamefont {Wettlaufer}}, \bibinfo {author}
  {\bibfnamefont {L.~A.}\ \bibnamefont {Wilen}}, \ and\ \bibinfo {author}
  {\bibfnamefont {E.~R.}\ \bibnamefont {Dufresne}},\ }\href {\doibase
  10.1103/PhysRevLett.110.066103} {\bibfield  {journal} {\bibinfo  {journal}
  {Phys. Rev. Lett.}\ }\textbf {\bibinfo {volume} {110}},\ \bibinfo {pages}
  {066103} (\bibinfo {year} {2013}{\natexlab{b}})}\BibitemShut {NoStop}%
\bibitem [{\citenamefont {Schulman}\ and\ \citenamefont
  {Dalnoki-Veress}(2015)}]{schulman2015}%
  \BibitemOpen
  \bibfield  {author} {\bibinfo {author} {\bibfnamefont {R.~D.}\ \bibnamefont
  {Schulman}}\ and\ \bibinfo {author} {\bibfnamefont {K.}~\bibnamefont
  {Dalnoki-Veress}},\ }\href@noop {} {\bibfield  {journal} {\bibinfo  {journal}
  {Physical review letters}\ }\textbf {\bibinfo {volume} {115}},\ \bibinfo
  {pages} {206101} (\bibinfo {year} {2015})}\BibitemShut {NoStop}%
\bibitem [{\citenamefont {Boltyanskiy}\ \emph {et~al.}(2017)\citenamefont
  {Boltyanskiy}, \citenamefont {Merrill},\ and\ \citenamefont
  {Dufresne}}]{Boltyanskiy2017}%
  \BibitemOpen
  \bibfield  {author} {\bibinfo {author} {\bibfnamefont {R.}~\bibnamefont
  {Boltyanskiy}}, \bibinfo {author} {\bibfnamefont {J.~W.}\ \bibnamefont
  {Merrill}}, \ and\ \bibinfo {author} {\bibfnamefont {E.~R.}\ \bibnamefont
  {Dufresne}},\ }\href {\doibase 10.1039/C6SM02011A} {\bibfield  {journal}
  {\bibinfo  {journal} {Soft Matter}\ }\textbf {\bibinfo {volume} {13}},\
  \bibinfo {pages} {2201} (\bibinfo {year} {2017})}\BibitemShut {NoStop}%
\bibitem [{\citenamefont {Style}\ \emph
  {et~al.}(2013{\natexlab{c}})\citenamefont {Style}, \citenamefont {Che},
  \citenamefont {Park}, \citenamefont {Weon}, \citenamefont {Je}, \citenamefont
  {Hyland}, \citenamefont {German}, \citenamefont {Power}, \citenamefont
  {Wilen}, \citenamefont {Wettlaufer},\ and\ \citenamefont
  {Dufresne}}]{Style2013_PNAS}%
  \BibitemOpen
  \bibfield  {author} {\bibinfo {author} {\bibfnamefont {R.~W.}\ \bibnamefont
  {Style}}, \bibinfo {author} {\bibfnamefont {Y.}~\bibnamefont {Che}}, \bibinfo
  {author} {\bibfnamefont {S.~J.}\ \bibnamefont {Park}}, \bibinfo {author}
  {\bibfnamefont {B.~M.}\ \bibnamefont {Weon}}, \bibinfo {author}
  {\bibfnamefont {J.~H.}\ \bibnamefont {Je}}, \bibinfo {author} {\bibfnamefont
  {C.}~\bibnamefont {Hyland}}, \bibinfo {author} {\bibfnamefont {G.~K.}\
  \bibnamefont {German}}, \bibinfo {author} {\bibfnamefont {M.~P.}\
  \bibnamefont {Power}}, \bibinfo {author} {\bibfnamefont {L.~A.}\ \bibnamefont
  {Wilen}}, \bibinfo {author} {\bibfnamefont {J.~S.}\ \bibnamefont
  {Wettlaufer}}, \ and\ \bibinfo {author} {\bibfnamefont {E.~R.}\ \bibnamefont
  {Dufresne}},\ }\href {\doibase 10.1073/pnas.1307122110} {\bibfield  {journal}
  {\bibinfo  {journal} {Proceedings of the National Academy of Sciences}\
  }\textbf {\bibinfo {volume} {110}},\ \bibinfo {pages} {12541} (\bibinfo
  {year} {2013}{\natexlab{c}})}\BibitemShut {NoStop}%
\bibitem [{\citenamefont {Olives}(2010)}]{olives2010}%
  \BibitemOpen
  \bibfield  {author} {\bibinfo {author} {\bibfnamefont {J.}~\bibnamefont
  {Olives}},\ }\href@noop {} {\bibfield  {journal} {\bibinfo  {journal} {J.
  Phys.: Condens. Matter}\ }\textbf {\bibinfo {volume} {22}},\ \bibinfo {pages}
  {085005} (\bibinfo {year} {2010})}\BibitemShut {NoStop}%
\bibitem [{\citenamefont {Hui}\ and\ \citenamefont {Jagota}(2013)}]{hui2013}%
  \BibitemOpen
  \bibfield  {author} {\bibinfo {author} {\bibfnamefont {C.-Y.}\ \bibnamefont
  {Hui}}\ and\ \bibinfo {author} {\bibfnamefont {A.}~\bibnamefont {Jagota}},\
  }\href@noop {} {\bibfield  {journal} {\bibinfo  {journal} {Langmuir}\
  }\textbf {\bibinfo {volume} {29}},\ \bibinfo {pages} {11310} (\bibinfo {year}
  {2013})}\BibitemShut {NoStop}%
\bibitem [{\citenamefont {Fuller}\ and\ \citenamefont
  {Vermant}(2012)}]{Fuller2012}%
  \BibitemOpen
  \bibfield  {author} {\bibinfo {author} {\bibfnamefont {G.~G.}\ \bibnamefont
  {Fuller}}\ and\ \bibinfo {author} {\bibfnamefont {J.}~\bibnamefont
  {Vermant}},\ }\href {\doibase 10.1146/annurev-chembioeng-061010-114202}
  {\bibfield  {journal} {\bibinfo  {journal} {Annual Review of Chemical and
  Biomolecular Engineering}\ }\textbf {\bibinfo {volume} {3}},\ \bibinfo
  {pages} {519} (\bibinfo {year} {2012})}\BibitemShut {NoStop}%
\bibitem [{\citenamefont {Pepicelli}\ \emph {et~al.}(2017)\citenamefont
  {Pepicelli}, \citenamefont {Verwijlen}, \citenamefont {Tervoort},\ and\
  \citenamefont {Vermant}}]{Pepicelli2017}%
  \BibitemOpen
  \bibfield  {author} {\bibinfo {author} {\bibfnamefont {M.}~\bibnamefont
  {Pepicelli}}, \bibinfo {author} {\bibfnamefont {T.}~\bibnamefont
  {Verwijlen}}, \bibinfo {author} {\bibfnamefont {T.~A.}\ \bibnamefont
  {Tervoort}}, \ and\ \bibinfo {author} {\bibfnamefont {J.}~\bibnamefont
  {Vermant}},\ }\href {\doibase 10.1039/C7SM01100H} {\bibfield  {journal}
  {\bibinfo  {journal} {Soft Matter}\ }\textbf {\bibinfo {volume} {13}},\
  \bibinfo {pages} {5977} (\bibinfo {year} {2017})}\BibitemShut {NoStop}%
\bibitem [{\citenamefont {Shenoy}(2005)}]{Shenoy2005}%
  \BibitemOpen
  \bibfield  {author} {\bibinfo {author} {\bibfnamefont {V.~B.}\ \bibnamefont
  {Shenoy}},\ }\href {\doibase 10.1103/PhysRevB.71.094104} {\bibfield
  {journal} {\bibinfo  {journal} {Phys. Rev. B}\ }\textbf {\bibinfo {volume}
  {71}},\ \bibinfo {pages} {094104} (\bibinfo {year} {2005})}\BibitemShut
  {NoStop}%
\end{thebibliography}

%

\end{document}